\documentclass[12pt]{article}
\usepackage{axodraw,bbold}

\catcode`@=12
\begin{document}
\thispagestyle{empty}
\begin{flushright} 
UCRHEP-T485\\ 
February 2010\
\end{flushright}
\vspace{0.5in}
\begin{center}
{\bf LEFT-HANDED NEUTRINOS AND RIGHT-HANDED SCOTINOS\\ 
(Secret Identity of the Right-handed Neutrino Exposed:\\ 
It is Actually a Scotino [Dark-Matter Fermion])\\}
\vspace{1.0in}
{\bf Ernest Ma\\}
\vspace{0.2in}
{\sl Department of Physics and Astronomy\\ 
University of California, Riverside\\ 
Riverside, California 92521, USA \\}
\vspace{1.0in}
\end{center}

\begin{abstract}\
The Standard Model of particle interactions is extended to include 
fermion doublets $(n,e)_R$ transforming under the gauge group $SU(2)_R$ 
such that $n$ is a scotino (dark-matter fermion), with odd $R$ parity. 
This dark left-right model (DLRM) treats neutrinos and scotinos in parallel, 
and has interesting phenomenology at the Large Hadron Collider (LHC).
\end{abstract}

\vspace{0.5in}

\noindent $^*$Talk at BUE-CTP International Conference on Neutrino Physics  
in the LHC Era, Luxor, Egypt, November 2009.

\newpage
\baselineskip 24pt
\section{Introduction}
In this talk, I will discuss the concept of having a ``right-handed'' singlet 
neutrino.  I will consider its role in the Standard Model of particle 
interactions, and in its left-right gauge extension. I will show how its 
true identity may be misread in the usual treatment of left-right models, 
and expose it as a ``scotino'', i.e. a dark-matter fermion. Interesting 
phenomenological and theoretical consequences of this hypothesis are 
presented.

\section{SM $\nu_R$ is not compulsory}
The Standard Model (SM) does not need a singlet fermion $\nu_R$ because it 
transforms as $(1,1,0)$, i.e. trivially under the gauge group 
$SU(3)_C \times SU(2)_L \times U(1)_Y$.  If $\nu_R$ is added anyway, 
then the Yukawa interaction $\bar{\nu}_R (\nu_L \phi^0 - l_L \phi^+)$ 
induces a Dirac mass term $m_D \bar{\nu}_R \nu_L$, as $\phi^0$ acquires a 
nonzero vacuum expectation value.  Thus $\nu_R$ is usually referred to as the 
``right-handed neutrino''.  However, since $\nu_R$ is a gauge singlet, 
it can have an arbitrary Majorana mass $m_R$.  The resulting $2 \times 2$ 
mass matrix is of the famous seesaw form
\begin{equation}
{\cal M}_\nu = \pmatrix{ 0 & m_D \cr m_D & m_R},
\end{equation}
which has the eigenvalues $m_R/2 \pm \sqrt{(m_R^2/4) + m_D^2}$.  Assuming 
$m_R >> m_D$ then implies $\nu_L$ is almost a mass eigenstate with mass 
eigenvalue $-m_D^2/m_R$.  This idea (1979) has tyrannized the thinking of 
neutrino mass for some 20 years.  On the other hand, it was certainly known 
already in the beginning that $\nu_R$ was not compulsory for neutrino mass 
in the SM, but it was not until about 10 years ago (1999) that other equally 
``natural'' mechanisms were also widely discussed.

In particular, a Higgs triplet $(\xi^{++},\xi^+,\xi^0)$ with a very small 
vacuum expectation value $\langle \xi^0 \rangle$ works just as well for 
$\nu_L$ to acquire a Majorana mass without any $\nu_R$.  However, it was 
incorrectly thought by the community at large for many years that this is 
somehow ``unnatural''.  To understand why this is also a seesaw mechanism 
(Type II) and just as natural as that (Type I) using $\nu_R$ with a large 
$m_R$, see for example the 1998 paper of Ma and Sarkar,\cite{ms98} where 
it is shown simply and explicitly that
\begin{equation}
V = m_\xi^2 \xi^\dagger \xi + \mu \xi^\dagger \Phi \Phi + ... ~\Rightarrow~
\langle \xi^0 \rangle \simeq {-\mu \langle \phi^0 \rangle^2 \over m_\xi^2},
\end{equation}
if $m_\xi^2$ is positive and large.

\section{Left-right $\nu_R$ is compulsory}
If the SM is extended to accommodate $SU(3)_C \times SU(2)_L \times SU(2)_R 
\times U(1)_X$, then the conventional assignment of
\begin{eqnarray}
(\nu,l)_L \sim (1,2,1,-1/2), &~& (\nu,l)_R \sim (1,1,2,-1/2), \\ 
(u,d)_L \sim (3,2,1,1/6), &~& (u,d)_R \sim (3,1,2,1/6),
\end{eqnarray}
implies the well-known result that $X = (B-L)/2$ and $Y = T_{3R} + (B-L)/2$. 
There must then be Higgs bidoublets
\begin{equation}
\phi = \pmatrix{\phi_1^0 & \phi_2^+ \cr \phi_1^- & \phi_2^0}, ~~~ 
\tilde{\Phi} = \pmatrix{\bar{\phi}_2^0 & -\phi_1^+ \cr -\phi_2^- & 
\bar{\phi}_1^0},
\end{equation}
both transforming as $(1,2,2,0)$, yielding lepton Dirac mass terms 
\begin{equation}
m_l = f_l \langle \phi_2^0 \rangle + f'_l \langle \bar{\phi}_1^0 \rangle, ~~~
m_\nu = f_l \langle \phi_1^0 \rangle + f'_l \langle \bar{\phi}_2^0 \rangle,
\end{equation}
and similarly in the quark sector.  This results in the appearance of 
phenomenologically undesirable tree-level flavor-changing neutral currents 
from Higgs exchange, as well as inevitable $W_L - W_R$ mixing.  If 
supersymmetry is imposed, then $\tilde{\Phi}$ can be eliminated, but then 
$({\cal M}_\nu)_{ij} \propto ({\cal M}_l)_{ij}$ as well as 
$({\cal M}_u)_{ij} \propto ({\cal M}_d)_{ij}$, contrary to what is observed. 
Hence the prevalent thinking is that $SU(2)_R \times U(1)_{B-L}$ is actually 
broken down to $U(1)_Y$ at a very high scale from an $SU(2)_R$ Higgs triplet 
$(\Delta_R^{++},\Delta_R^+,\Delta_R^0) \sim (1,1,3,1)$ which provides $\nu_R$ at 
the same time with a large Majorana mass from $\langle \Delta_R^0 \rangle$. 

The Type I seesaw mechanism is thus implemented and everyone should be happy. 
But wait, no remnant of the $SU(2)_R$ gauge symmetry is detectable at the 
TeV scale and we will not know if $\nu_R$ really exists.  Is there a natural 
way to lower the $SU(2)_R \times U(1)_{B-L}$ breaking scale?

The answer was already provided 22 years ago\cite{m87} in the context of 
the superstring-inspired supersymmetric $E_6$ model.  The fundamental 
\underline{27} fermion representation here is decomposed under 
$[(SO(10),SU(5)]$ as
\begin{equation}
\underline{27} = (16,10) + (16,5^*) + (16,1) + (10,5) + (10,5^*) + (1,1).
\end{equation}
Under its maximum subgroup $SU(3)_C \times SU(3)_L \times SU(3)_R$, the 
\underline{27} is organized instead as $(3,3^*,1)+(1,3,3^*)+(3^*,1,3)$, i.e.
\begin{equation}
\pmatrix{d & u & h \cr d & u & h \cr d & u & h} + \pmatrix{N & E^c & \nu \cr 
E & N^c & e \cr \nu^c & e^c & n^c} + \pmatrix{d^c & d^c & d^c \cr u^c & u^c 
& u^c \cr h^c & h^c & h^c}.
\end{equation}
It was realized\cite{m87} in 1987 that there are actually two left-right 
options: (A) Let $E_6$ break down to the fermion content of the conventional 
$SO(10)$, given by $(16,10)+(16,5^*)+(16,1)$, which is the usual 
left-right model which everybody knows. (B) Let $E_6$ break down to the 
fermion content given by $(16,10)+(10,5^*)+(1,1)$ instead, thereby switching 
the first and third rows of $(3^*,1,3)$ and the first and third columns of 
$(1,3,3^*)$. Thus $(\nu,e)_R$ becomes $(n,e)_R$ and $n_R$ is not the mass 
partner of $\nu_L$. This is referred to by the Particle Data Group as the 
Alternative Left-Right Model (ALRM).  Here the usual left-handed lepton 
doublet is part of a bidoublet:
\begin{equation}
\pmatrix{\nu & E^c \cr e & N^c}_L \sim (1,2,2,0).
\end{equation}
In this supersymmetric model, $\nu_L$ is still the Dirac mass partner of 
$\nu_R$ and gets a seesaw mass, whereas\cite{m00} $n_R$ (which couples to 
$e_R$ through $W_R$) mixes with the usual neutralinos, the lightest of which 
is a dark-matter candidate.

\section{Dark left-right model}
Earlier in 2009, a simpler nonsupersymmetric variant of the ALRM was 
proposed\cite{klm09} which has the same basic fermion structure as a 
model discussed already 31 years ago\cite{rr78}.  We call it the Dark 
Left-Right Model (DLRM).  We impose a global U(1) symmetry $S$, so that 
under $SU(3)_C \times SU(2)_L \times SU(2)_R \times U(1) \times S$, the 
``leptons'' transform as
\begin{equation}
\psi_L = (\nu,e)_L \sim (1,2,1,-1/2;1), ~~~  
\psi_R = (n,e)_R \sim (1,1,2,-1/2;1/2),
\end{equation}
and the Higgs bidoublet as
\begin{equation}
\Phi = \pmatrix{\phi_1^0 & \phi_2^+ \cr \phi_1^- & \phi_2^0} \sim (1,2,2,0;1/2).
\end{equation}
Hence $\tilde{\Phi}$ has $S=-1/2$ and the Yukawa term $\bar{\psi}_L 
\tilde{\Phi} \psi_R$ is forbidden, whereas $\bar{\psi}_L \Phi \psi_R$
is allowed.  The breaking of $SU(2)_R \times U(1) \to U(1)_Y$ leaves $L = S - 
T_{3R}$ unbroken, so that $\langle \phi_2^0 \rangle \neq 0$, but $\langle 
\phi_1^0 \rangle = 0$.  The former allows a Dirac mass term $m_e \bar{e}_L 
e_R$, whereas the latter means that $\nu_L$ and $n_R$ are not Dirac mass 
partners and can be completely different particles with independent masses 
of their own.  Since $n_R$ has $L = 1/2 - 1/2 = 0$, it also has odd $R$ 
parity, i.e. $R = (-)^{3B+L+2j} = -1$, even though the model is 
nonsupersymmetric.  It may thus be a dark-matter fermion, i.e. a scotino.

Let $n_R$ and $\nu_L$ become massive in parallel, the former {\it via} 
$(\Delta_R^{++},\Delta_R^+,\Delta_R^0) \sim (1,1,3,1;-1)$, and the latter 
{\it via} $(\Delta_L^{++},\Delta_L^+,\Delta_L^0) \sim 
(1,3,1,1;-2)$.  Since $\Delta_L^0$ has $L=-2$, the soft term 
$\tilde{\Phi}_L^\dagger \Delta_L \Phi_L$ is needed to break $L$ to $(-)^L$. 
The Higgs doublet $\Phi_L \sim (1,2,1,1/2;0)$ is needed as well as 
$\Phi_R \sim (1,1,2,1/2;-1/2)$ because the quark sector is now given by 
\begin{eqnarray}
&& Q_L = (u,d)_L \sim (3,2,1,1/6;0), ~~~ d_R \sim (3,1,1,-1/3;0), \\ 
&& Q_R = (u,h)_R \sim (3,1,2,1/6;1/2), ~~~ h_L \sim (3,1,1,-1/3;1).
\end{eqnarray}
The allowed Yukawa terms are then $\bar{Q}_L \tilde{\Phi} Q_R$, $\bar{Q}_L 
\Phi_L d_R$, and $\bar{Q}_R \Phi_R h_L$.  Hence $m_u$ comes from $v_2 = 
\langle \phi_2^0 \rangle$, $m_d$ from $v_3 = \langle \phi_L^0 \rangle$, and 
$m_h$ from $v_4 = \langle \phi_R^0 \rangle$.  Tree-level flavor-changing 
neutral currents are thus guaranteed to be absent.  Since the scotino $n$ has 
$L=0$ and $e$ has $L=1$, this implies that $W_R^+$ has $L=-1$ and $h$ has 
$L=1$. Thus $W_R$ does not mix with $W_L$, and $h$ does not mix with $d$.

The new neutral gauge boson $Z'$ of this model couples to the current 
$J_{Z'} = x J_{3L} + (1-x) J_{3R} - x J_{em}$ where $x \equiv \sin^2 \theta_W$, 
with coupling $e/\sqrt{x(1-x)(1-2x)}$.  Neglecting $M_{W_L}^2$, we then have
\begin{equation}
{(1-2x) \over 2(1-x)} M_{Z'}^2 < M_{W_R}^2 < {(1-2x) \over (1-x)} M_{Z'}^2.
\end{equation}
The lower (upper) bound applies to $\langle \phi_R^0 \rangle << (>>) 
\langle \Delta_R^0 \rangle$.  Present Tevatron data imply that 
$M_{Z'} > 850$ GeV ($M_{W_R} > 500$ GeV).  At the LHC (with $E_{cm}=14$ TeV), 
the discovery reach\cite{klm09} of this $Z'$ by the observation of 10 
dilepton events of one type is $M_{Z'} = 1.5$ (2.4) TeV for an integrated 
luminosity of 1 (10) fb$^{-1}$.  Once $Z'$ is observed, this model predicts 
$W_R$ as shown, in contrast to all purely $U(1)'$ gauge models.

\section{Scotino = $n_R$ ($\nu_R$)}
The particles $n,h,W_R^\pm,\phi_R^\pm,\Delta_R^\pm,\phi_1^\pm,Re(\phi_1^0),
Im(\phi_1^0)$ are odd under $R$ parity.  The lightest $n$ can be stable and 
be a good candidate for the dark matter of the Universe.  Assuming that 
$\Delta_R^\pm$ is much lighter than $W_R^\pm$ and $Z'$, the dominant 
annihilation of $n$ is then $n n \to e^+ e^-$ {\it via} $\Delta_R^\pm$ 
exchange.  The measured $\Omega h^2$ values for dark matter by WMAP are 
obtained\cite{klm09} for a wide range of $n$ and $\Delta_R^\pm$ masses 
in the neighborhood of 200 GeV.  Since $n$ always interacts with a lepton 
in this model, recent observations by the PAMELA and ATIC collaborations 
may also be relevant.\cite{cms09}

\section{Scotogenic neutrino masses}
The mass of $\nu_L$ may also be derived from $n_R$ as a radiative 
effect, i.e. scotogenic.  To accomplish this, $\Delta_L$ is removed in 
favor of of a scalar singlet $\chi \sim (1,1,1,0;-1)$, then the trilinear 
scalar term Tr$(\Phi \tilde{\Phi}^\dagger)\chi$ is allowed.  Using the 
soft term $\chi^2$ to break $L$ to $(-)^L$, a scotogenic  neutrino mass 
is obtained in one loop, as pointed out first\cite{m06} in 2006.

\section{Conclusion}
The presence of $\nu_R$ is unavoidable in a left-right gauge extension of the 
Standard Model.  However, it does not have to be the Dirac mass partner of 
$\nu_L$.  In that case, it should be renamed $n_R$ and could function as a 
scotino, i.e. a dark-matter fermion.  The $SU(2)_R$ gauge bosons of this 
dark left-right model (DLRM), i.e. $W_R^\pm$ and $Z'$, are observable at the 
LHC. The recent PAMELA and ATIC observations may also be relevant. 
Scotogenic neutrino masses are also possible.

\section*{Acknowledgments} 
This work was supported in part by the 
U.~S.~Department of Energy under Grant No.~DE-FG03-94ER40837.  I thank 
S. Khalil and the other organizers of the BUE-CTP International Conference 
on Neutrino Physics in the LHC Era (November 2009) for their great 
hospitality and a stimulating meeting in Luxor.

\newpage 

\begin{thebibliography}{9}   

\bibitem{ms98} E. Ma and U. Sarkar, Phys. Rev. Lett. {\bf 80}, 5716 (1998).
\bibitem{m87} E. Ma, Phys. Rev. {\bf D36}, 274 (1987).
\bibitem{m00} E. Ma, Phys. Rev. {\bf D62}, 093022 (2000).
\bibitem{klm09} S. Khalil, H.-S. Lee, and E. Ma, Phys. Rev. {\bf D79}, 
041701(R) (2009).
\bibitem{rr78} P. Ramond and D. B. Reiss, Phys. Lett. {\bf 80B}, 87 (1978).
\bibitem{cms09} Q.-H. Cao, E. Ma, and G. Shaughnessy, Phys. Lett. {\bf B673}, 
152 (2009).
\bibitem{m06} E. Ma, Phys. Rev. {\bf D73}, 077301 (2006).
\end{thebibliography}

\end{document}